%% file: main.tex
\documentclass[11pt,a4paper]{article}

\usepackage[utf8]{inputenc}
\usepackage[T1]{fontenc}
\usepackage[margin=2.5cm]{geometry}
\usepackage{graphicx}
\usepackage{booktabs}
\usepackage{tabularx}
\usepackage{subcaption}
\usepackage{amsmath, amsthm, amsfonts, amssymb}
\usepackage{braket}
\usepackage[colorlinks=true, allcolors=blue]{hyperref}

\providecommand{\abs}[1]{\lvert#1\rvert}
\providecommand{\ketbra}[2]{\ket{#1}\bra{#2}}

\title{Quantum iterative approach to the Traveling Salesman Problem}

\author{
  Arturo Rodríguez-Almazán$^{1,*}$,
  Guillermo Rivas$^{2}$,
  Daniela Falcó$^{2}$,\\
  Mir Amir Hosseini$^{2}$,
  and Ricardo S. Alonso$^{1,3}$\\[0.75em]
  \footnotesize $^{1}$AIR Institute, Carr. de la Aldehuela, 49022, Zamora, Castilla y León, Spain\\
  \footnotesize $^{2}$BISITE Research Group, University of Salamanca, Calle Espejo, 37007, Salamanca, Castilla y León, Spain\\
  \footnotesize $^{3}$International University of La Rioja, Av. de la Paz, 26006, Logroño, La Rioja, Spain\\[0.5em]
  \footnotesize $^{*}$Correspondence: \href{mailto:arodrigueza@air-institute.es}{arodrigueza@air-institute.es}
}

\date{}

\begin{document}

\noindent\fbox{%
  \begin{minipage}{\dimexpr\textwidth-2\fboxsep-2\fboxrule\relax}
    \small
    \textbf{Postprint / Accepted Manuscript Notice:} Accepted for publication in \textit{Physica Scripta}. This is the Accepted Manuscript version of an article published by IOP Publishing in \textit{Physica Scripta}, vol.~101, 425101 (2026). The Version of Record is available online at \url{https://doi.org/10.1088/1402-4896/aeb04d}.\\[0.4em]
    \textbf{Citation:} Arturo Rodríguez-Almazán \textit{et al.} 2026 \textit{Phys. Scr.} \textbf{101} 425101\\
    \textbf{Journal-ref:} Phys. Scr. 101 425101 (2026)\\
    \textbf{DOI:} \href{https://doi.org/10.1088/1402-4896/aeb04d}{10.1088/1402-4896/aeb04d}
  \end{minipage}%
}

\vspace{1em}

{\let\newpage\relax\maketitle}

\begin{abstract}
\input{abstract}
\end{abstract}

\vspace{0.75em}

\section{Introduction}
\input{Introduction}

\section{Phase Estimation}\label{seccion pe}
\input{phaseestimation}

\section{Amplitude Amplification}\label{amplitudamplification sec}
\input{amplitudeamplification}

\section{The TSP iterative solver}\label{tspsolver}
\input{tspsolver}

\section{Use case}\label{Usecase}
\input{practicalcase}

\section{Conclusions and future work}\label{conclusion}
\input{conclusion}

\section*{Funding}
This work has been funded by the Centro para el Desarrollo Tecnológico y la Innovación (CDTI) under the Cervera Program--Network of Excellence, promoted by the Spanish Ministry of Science and Innovation through the project `Alianza de centros tecnológicos para el Quantum Computing Accesible y dual en Defensa y sectores Estratégicos (ARQADE)'' (grant number~CER-20251019).

\section*{Supplementary data}
\input{matematicasgroverlong}

\bibliographystyle{unsrt}
\bibliography{BiblioTSP}

\end{document}

%% file: abstract.tex
The Traveling Salesman Problem remains a fundamental NP-hard challenge in combinatorial optimization, where the factorial growth of the search space renders exact classical solutions computationally prohibitive as the problem size increases. To address this limitation, this work explores digital quantum computing as an alternative paradigm and proposes a hybrid, iterative framework that integrates Quantum Phase Estimation and Quantum Amplitude Amplification implemented via the Grover-Long algorithm. Within this architecture, normalized route costs are encoded as quantum phases, while Amplitude Amplification combined with an external classical loop progressively updates a cost threshold to filter and refine the solution space toward the optimal tour. The operational feasibility of the framework is validated through a proof-of-concept case study simulating both symmetric and asymmetric four-city TSP instances, demonstrating successful convergence. Furthermore, under an expectation-based runtime analysis, the algorithm exhibits an expected computational complexity of $\mathcal{O}\left(m^{5/2}\log^2(m)\right)$. Critically, this scaling evaluation treats the state preparation component within the iterative loop as subdominant to the Phase Estimation stage. However, it must be emphasized that the current baseline approach implements this state preparation inefficiently, scaling factorially; thus, the overall quantum advantage hinges entirely on resolving this bottleneck. While the desired state possesses a highly defined mathematical structure that suggests potential for optimization, suppressing this overhead remains a strict prerequisite.

%% file: Introduction.tex
The Traveling Salesman Problem (TSP)~\cite{flood_traveling-salesman_1956} is one of the most fundamental and well-studied problems in combinatorial optimization, with significant applications in logistics, manufacturing, and network routing. It involves finding the shortest possible route that visits a given set of cities exactly once and returns to the starting point, minimizing the total travel cost. In other words, the goal is to determine an optimal tour that covers all locations without repetition and completes the circuit efficiently. Formally, the problem can be modeled as a graph $G = (V, E)$, where vertices $V$ represent the cities and edges $E$ represent the paths between them, each associated with a specific cost. The objective is then reduced to finding a Hamiltonian cycle with the minimum total weight. Despite its simple formulation, the problem is computationally challenging due to its NP-hard nature. As the number of cities $m$ increases, the number of unique possible routes grows factorially, specifically as $(m-1)!$, leading to an exponential explosion of the search space, which makes brute-force enumeration completely unfeasible for classical computers~\cite{garey2002computers}. Consequently, obtaining exact solutions for large-scale instances within reasonable time frames becomes mathematically intractable. This limitation has historically forced classical frameworks to rely on approximation algorithms and metaheuristics. However, while these methods successfully trade absolute optimality for computational feasibility, they often fail to guarantee global upper bounds as the problem size expands. This complexity has positioned TSP as a benchmark for evaluating optimization techniques and algorithmic paradigms, both in classical and quantum computational models.

In the remainder of this work, we will investigate the application of digital quantum computing techniques to address the Traveling Salesman Problem. Specifically, we present an approach that builds upon Quantum Phase Estimation~\cite{kitaev1995quantum} and  Quantum Amplitude Amplification~\cite{brassard2002quantum}, an expansion of Grover's search algorithm~\cite{grover1997quantum}. Rather than relying on a unique quantum circuit, our architecture utilizes a classical external loop that drives the execution of the quantum circuits across successive iterations. Within this hybrid loop, Quantum Phase Estimation is utilized to compute and encode route costs, effectively serving as the state initializer for the Amplitude Amplification operator. Subsequently, Amplitude Amplification acts as a filter that targets only the states falling below a specific cost threshold. By iteratively lowering this threshold through the classical loop, the architecture progressively narrows the solution space until the absolute minimum tour is isolated with high probability. Before presenting the proposed methodology, Section~\ref{relatedwork} reviews related work, including classical approaches and recent quantum strategies for solving the TSP. Subsequently, Sections~\ref{seccion pe} and \ref{amplitudamplification sec} introduce the theoretical foundations of both algorithms, Section~\ref{tspsolver} describes the TSP iterative solver, and Section~\ref{Usecase} illustrates its operation through both symmetric and asymmetric practical use cases, followed by our concluding remarks in Section~\ref{conclusion}.

\section{Related work}\label{relatedwork}

Over the years, a wide range of classical techniques have been developed to address the Traveling Salesman Problem, achieving significant progress for small and medium instances. Among the most notable approaches are branch-and-bound~\cite{balas1983branch}, branch-and-cut implementations~\cite{applegate1998solution} and dynamic programming~\cite{bellman_dynamic_1962, held1962dynamic}. In particular, the Concorde TSP Solver, based on branch-and-cut, represents the state of the art for exact solutions, having solved instances with more than eight thousand cities~\cite{applegate2011traveling}. When exact methods become computationally prohibitive, heuristic and metaheuristic algorithms are employed. Traditional heuristics~\cite{lin1973effective} have been complemented by nature-inspired metaheuristics, such as Ant Colony Optimization~\cite{dorigo2002ant} and Genetic Algorithms~\cite{ulder1990genetic}. In recent years, the classical state of the art has shifted towards modifying and improving metaheuristic and swarm intelligence algorithms. Researchers have developed models specifically tailored for the TSP such as the Farmland Fertility Algorithm (FFA) enhanced with hyper-heuristic approaches~\cite{gharehchopogh2023improved} or integrated with the Metropolis acceptance criterion~\cite{abdollahzadeh2021discrete}. Similarly, specialized versions of the Harris Hawk Optimization algorithm~\cite{gharehchopogh2022efficient} have been successfully adapted to handle the discrete nature of the TSP. However, despite these continuous algorithmic refinements, classical scalability remains limited, and exact or high-quality solutions for large-scale TSP instances still suffer from the exponential growth of the solution space. 

In recent years, quantum computing has emerged as a promising paradigm for tackling NP-hard problems, including TSP. Most existing research in this area has focused on quantum annealing~\cite{martovnak2004quantum, warren2013adapting}, where TSP is encoded as a Quadratic Unconstrained Binary Optimization  (QUBO) problem~\cite{lucas2014ising}. This formulation allows leveraging specialized hardware, such as quantum annealers, to approximate solutions by exploiting quantum tunneling and energy minimization. The natural transition from this analog paradigm to gate-based quantum computing is therefore articulated through variational algorithms. Specifically, the Quantum Approximate Optimization Algorithm (QAOA)~\cite{farhi2014quantum} has become the benchmark for solving QUBO TSP instances on digital quantum devices~\cite{ruan2020quantum}. While these QUBO-based studies have demonstrated the feasibility of quantum approaches for combinatorial optimization, the exploration of digital quantum algorithms for TSP remains relatively limited. Other notable contributions include Grover-based algorithms such as~\cite{zhu2022realizable}, as well as the approach presented in~\cite{goswami2024solving}, which introduces a novel strategy using just one qubit. Among these related works,~\cite{srinivasan2018efficient} is of particular importance as it provides the cost calculation method that serves as the foundation for the approach proposed in this work. A comparison between spatial and time complexity of these methods is presented in Table~\ref{tablacomplx}. As indicated by this comparative overview, existing digital quantum approaches are limited and do not yet outperform classical approaches. To address these limitations, our framework shifts toward the quantum-classical iterative TSP solver. Consequently, before detailing the integrated iterative solver, it is essential to establish the mathematical and theoretical foundations of its core quantum building blocks. We begin by reviewing the Quantum Phase Estimation algorithm, which serves as the primary mechanism for evaluating route costs within our system.

\begin{table}[h!]
    \centering
    \caption{Time and space complexity of TSP algorithms}
    \vspace{0.2cm}
    \renewcommand{\arraystretch}{1.4}
    \begin{tabularx}{\textwidth}{@{}X l l l@{}}
        \toprule
        \textbf{Reference} & \textbf{Method} & \textbf{Time} & \textbf{Space} \\
        \midrule
        \multicolumn{4}{l}{\textbf{Classical Exact Methods}} \\
        \midrule
        Balas \& Toth (1983) \cite{balas1983branch}& Branch and Bound & $\mathcal{O}(m^3)\to\mathcal{O}(m!)$ & $\mathcal{O}(m^2)$  \\
        Applegate et al. (1998) \cite{applegate1998solution}& Cutting Planes & $\mathcal{O}(m^3)\to\mathcal{O}(m!)$ & $\mathcal{O}( 2^m)$ \\
         Held \& Karp (1962) \cite{held1962dynamic} & Dynamic Programming & $\mathcal{O}(m^2 \cdot 2^m)$ & $\mathcal{O}(m \cdot 2^m)$ \\
        \midrule
        \multicolumn{4}{l}{\textbf{Classical Heuristics and Metaheuristics}} \\
        \midrule
        Lin \& Kernighan (1973) \cite{lin1973effective} & Local Search & $\mathcal{O}(m^{2.2})$ (Empirical) & $\mathcal{O}(m^2)$ \\
        Dorigo \& Gambardella (1997) \cite{dorigo2002ant} & Ant Colony System  & $\mathcal{O}(I \cdot P \cdot m)$ & $\mathcal{O}(m^2)$ \\
        Ulder  et al. (1990) \cite{ulder1990genetic} & Genetic Local Search & $\mathcal{O}(I \cdot P \cdot m^2)$ & $\mathcal{O}(P \cdot m)$ \\
        Gharehchopogh et al. (2023) \cite{gharehchopogh2023improved}& FFA  &  $\mathcal{O}(I \cdot P \cdot m^2)$ & $\mathcal{O}(P \cdot m)$ \\
        \midrule
        \multicolumn{4}{l}{\textbf{Quantum Algorithms}} \\
        \midrule
        Warren (2013) \cite{warren2013adapting}& Quantum Annealing & \multicolumn{1}{c}{$\sim$} & $\mathcal{O}(m^2)$ \\
        Ruan et al. (2020) \cite{ruan2020quantum}& QAOA & $\mathcal{O}(I\cdot p \cdot m^3)$ & $\mathcal{O}(m^2)$\\
         Zhu et al. (2022) \cite{zhu2022realizable}& Grover-based & $\mathcal{O}(2^{Dm/2}\cdot m^2\log^2(m))$& $\mathcal{O}(D\cdot m)$\\
        Goswami et al. (2024) \cite{goswami2024solving} & Single Qubit Heuristic & Exp. Sampling & $1$ qubit \\
        \bottomrule
    \end{tabularx}\\
    \vspace{0.15 cm}
    \footnotesize
    \noindent \textbf{Nomenclature:} 
    $I$: Number of iterations, generations, or cycles; 
    $P$: Population size; 
    $p$: Quantum layers in QAOA; 
    $D$: Minimum bits required to store the maximum distance.
    \label{tablacomplx}
\end{table}

%% file: phaseestimation.tex
\begin{figure}[t]
\centering
\includegraphics[width=0.65\textwidth]{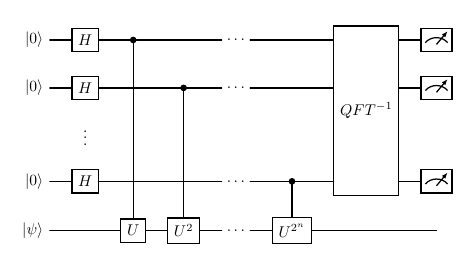}
\caption{Schematic diagram representing the quantum circuit to perform Quantum Phase Estimation.}
\label{QPE}
\end{figure}

The Quantum Phase Estimation (QPE)~\cite{kitaev1995quantum} algorithm is one of the cornerstone quantum algorithms. It enables quantum computers to estimate the eigenvalue (or phase) corresponding to an eigenstate of an operator. It is, for example, the key element in Shor's famous algorithm for integer factorization~\cite{shor_algorithms_1994}. Given an operator $U$ and a state $\ket{\psi}$ that is an eigenstate of $U$, the objective of Phase Estimation is to find the value $\phi$ such that:
\begin{equation}
    U \ket{\psi}= e^{2i \pi \phi}\ket{\psi}.
    \label{eigenstate}
\end{equation}
The QPE algorithm (Figure~\ref{QPE}) commences with the initialization of two quantum registers: a control register comprising $n$ qubits set to $\ket{0}$ and a target register prepared in the eigenstate 
$\ket{\psi}$. The algorithm applies a Hadamard gate to each qubit in the control register, thereby creating a superposition of all possible states. Subsequently, the controlled application of the unitary $U$ is performed, where each control qubit dictates the application of $U$ raised to successive powers of two ($U^{2^k}$), encoding the phase information of the eigenvalues into the control qubits by the phase kick-back phenomenon~\cite{nielsen2010quantum}. Following the completion of this encoding process, an inverse Quantum Fourier Transform~\cite{coppersmith2002approximate} is applied to the control register, thereby transforming the superposition into a state in which the binary representation of the phase $\phi$ is accessible. Finally, a measurement of the control register yields an approximation of $\phi$. The precision of this approximation is determined by the number of control qubits utilized.

A concise mathematical exposition of this algorithm will now be provided. After the Hadamard application, the system is in the state:
\begin{equation}
    \ket{\Psi_0}=\frac{1}{2^{\frac{n}{2}}}\sum_{x=0}^{2^n-1}\ket{x} \ket{\psi}.
\end{equation}
We can notice that all the controlled gates can be interpreted as the operator:
\begin{equation}
    CU_T\ket{x} \ket{\psi}=\ket{x} U^x \ket{\psi} = e^{2i \pi x \phi}\ket{x} \ket{\psi}.
    \label{phase est}
\end{equation}
Where the last equality is trivial given Eq.~\ref{eigenstate}. We can apply this operator to the $\ket{\Psi_0}$ state obtaining:
\begin{equation}
    \ket{\Psi_1}=\frac{1}{2^{\frac{n}{2}}}\sum_{x=0}^{2^n-1} e^{2i \pi x \phi} \ket{x} \ket{\psi}.
\end{equation}
Now  the inverse QFT is applied to the control register, resulting in:
\begin{equation}
    \ket{\Psi_2}=\frac{1}{2^n} \sum_{x'=0}^{2^n-1} \sum_{x=0}^{2^n-1} e^{2i \pi x \phi} e^{-\frac{2i \pi}{2^n} x x'} \ket{x'} \ket{\psi}.
\end{equation}
This expression can be rewritten in a more appropriate form:
\begin{equation}
    \ket{\Psi_2}=\frac{1}{2^n} \sum_{x'=0}^{2^n-1} \sum_{x=0}^{2^n-1}  e^{-\frac{2i \pi x}{2^n} (x'-2^n \phi)} \ket{x'} \ket{\psi}.
\end{equation}
As it can be observed in the final equation, when we measure the control register,  the probability is maximized at $x' \approx 2^n \phi $. However, as $x'$ deviates from $2^n \phi$, a destructive interference emerges, and the probability tends to zero. Notably, when $x' = 2^n \phi $, the probability of measuring $ 2^n \phi $ in the control register is exactly 1, meaning that the algorithm succeeds with a $100$ \% probability. Conversely, if $2^n \phi$ is not an integer, the phase cannot be represented perfectly using $n$ qubits, causing the success probability to strictly drop below $100\%$ and leaving a non-zero chance of failure. This algorithm facilitates the identification of the phase associated with a specific operator. In the context of our TSP solver, this operator is employed to encode the cost of each different route, which is introduced as each distinct eigenstate $\ket{\psi}$. Consequently, after the application of QPE the result in the control register $\ket{x'}$ represents the cost value of the route introduced as eigenstate.

%% file: amplitudeamplification.tex
\begin{figure}[ht]
\centering
\includegraphics[width=0.65\textwidth]{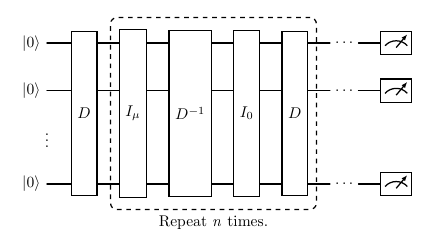}

\caption{Schematic diagram representing the quantum circuit to perform the Grover-Long algorithm.}
\label{Am mplitud A circuit}
\end{figure}

Another pivotal algorithm for quantum computers that will be utilized is Amplitude Amplification. This was first proposed by Grover~\cite{grover1997quantum,grover_fast_1996}, and its primary objective is to enhance the probability of achieving specific states given an initial superposition. It operates by repeatedly applying two key operators: the oracle and the amplification operator. The oracle marks the target states ($\ket{\mu}$) by applying a phase flip. The amplification operator then amplifies the amplitude of the marked states by reflecting the quantum state around the average amplitude of all states. In this study, we employ a modified version known as the Grover-Long algorithm~\cite{long2001grover} (Figure~\ref{Am mplitud A circuit}). This modification involves the execution of distinct phase rotations instead of just phase flips.

As we have done for Phase Estimation, we will now give a short mathematical description. Firstly, we will examine the two key operators, the oracle and the amplification operator. We need the oracle to introduce a $\phi$ phase shift in the desired states; to that end, $I_\mu=I+(e^{i \phi}-1)\ket{\mu} \bra{\mu}$ will be used. For the second operator we extend the definition given in~\cite{grover1998quantum}, taking into account that as in the oracle, we are introducing angle rotations instead of the mean inversion. Therefore,  the operator is constructed as $I_\psi=I+(e^{i \phi}-1)\ket{\psi} \bra{\psi} =DI_0D^{-1}$, where $D$ represents the quantum circuit used for initialize the register to the initial superposition ($D\ket{0}=\ket{\psi}$) and $I_0=I+(e^{i \phi}-1)\ket{0} \bra{0} $ introduces the $\phi$ phase shift to the zero state. This means that in each Grover iteration, we apply $Q=-DI_0D^{-1}I_\mu$. In order to show how the operator $Q$ affects our state we will separate $\ket{\psi}$ in the ``good'' $\ket{\mu}$ and the ``bad'' $\ket{\tau}$ states as $\ket{\psi}= \sin{\beta} \ket{\mu} + \cos{\beta} \ket{\tau}$ where $\sin\beta= \sqrt{\frac{N'}{N}} $  denotes the number of solutions that exists within the total number of states. Now, it is illuminating to take a look at the solution of applying the $Q$ operator to both $\ket{\tau}$ and $\ket{\mu}$:
\begin{equation}
    Q\ket{\tau}=-I_\psi \ket{\tau}=-\ket{\tau}-(e^{i \phi}-1)\ket{\psi}\bra{\psi}\tau \rangle.
\end{equation}
\begin{equation}
    Q\ket{\mu}=-I_\psi e^{i \phi} \ket{\mu}=-e^{i \phi} \ket{\mu}-e^{i \phi} (e^{i \phi}-1)\ket{\psi}\bra{\psi}\mu \rangle.
\end{equation}
From the definition of $\ket{\psi}$ we have that $\ket{\psi}\bra{\psi}\tau \rangle= \sin{\beta} \cos{\beta} \ket{\mu} + \cos^2{\beta} \ket{\tau}$ and $\ket{\psi}\bra{\psi}\mu \rangle= \sin^2{\beta} \ket{\mu} +\sin{\beta} \cos{\beta} \ket{\tau}$. Introducing this, and after a few algebraic transformations, we can express $Q$ as a matrix in the $\braket{\ket{\mu},\ket{\tau}}$ basis:
\begin{equation}
    Q=\begin{bmatrix} -e^{i \phi} (1+(e^{i \phi}-1)\sin^2{\beta}) &-(e^{i \phi}-1) \sin{\beta} \cos{\beta}  \\-e^{i \phi} (e^{i \phi}-1) \sin{\beta} \cos{\beta}  & -e^{i \phi} + (e^{i \phi}-1)\sin^2{\beta}  \end{bmatrix} .
\end{equation}
As shown in Supplementary data, after $n\ge \frac{\frac{\pi}{2}-\beta}{2\beta}$ iterations of the $Q$ operator with:
\begin{equation}
    \phi= 2 \arcsin\bigg(\frac{\sin(\frac{\pi}{4n+2})}{\sin(\beta)} \bigg),
\end{equation}
we will find ourselves in the $\ket{\mu}$ state with probability equal 1. Note that this $100 \%$ probability of success is achieved under the assumption that the exact relationship between the number of solutions and the total number of states, characterized by the parameter $\beta$, is known a priori. In our TSP solver, we will use Amplitude Amplification for minimum finding, as was first proposed in~\cite{durr1996quantum}. This entails marking the states lower than a certain  value, then measuring randomly one of those states, and rerunning the algorithm using the measured value as a new threshold. This process is repeated until the minimum is reached.

%% file: tspsolver.tex
Once we have gained an insight into the two key elements, we provide an intuitive overview of the complete algorithm. As we have just mentioned, Grover-Long algorithm will be used as a minimum finder; to that end, we need to construct states of the form $\ket{Cost}\otimes \ket{Route}$, after this point they will be represented as $\ket{C}\otimes \ket{R}$. Consequently, if we reach the state with minimum value in the $\ket{C}$ register, we can measure the route which leads to that minimum cost in the $\ket{R}$ register. The Phase Estimation algorithm will be used to construct such states. 

Before going deeper into the algorithm, it is important to focus on the route encoding we will use, given in~\cite{srinivasan2018efficient}. With this encoding we need $m\lceil\log_2(m)\rceil$ qubits for the $\ket{R}$ register where $\lceil \log_2(m) \rceil$ represents the first integer greater than or equal to $\log_2(m)$. The $\ket{R}$ register is separated into groups of $\lceil \log_2(m) \rceil$ qubits, each group representing a city based on the position it takes along the register. The binary value of each of these groups indicates the city from which we arrive at the city represented. In order to clarify it, we can present an example: Assuming four cities, $\lceil \log_2(m) \rceil=2$ meaning that the $\ket{R}$ register is formed by eight qubits (four groups of two), the route $ 00 \to 11 \to 01 \to10\to 00$  will be $\ket{10}\otimes\ket{11}\otimes\ket{01}\otimes\ket{00} = \ket{10110100}$. For the $\ket{C}$ register the number of bits ($d$) determines the resolution required to distinguish between different route costs without losing precision, the error $\epsilon = 2^{-d}$ represents the smallest cost difference the register can resolve. Assuming individual path costs are acquired with a fixed measurement precision, the minimum possible difference between any two distinct routes is strictly bounded by this initial discrete resolution. Since all route values are normalized within the $[0, 1]$ range by dividing the individual paths by $m$ times the longest individual path, aggregating values across $m$ stages, digital accumulators experience a well known bit growth effect. To preserve the original resolution of the data across the entire tour, the register's bit depth must scale logarithmically with the number of cities $\mathcal{O}(\log_2 (m))$~\cite{oppenheim1972effects}. This logarithmic scaling is the theoretical minimum required to accommodate the sum of $m$ fixed-precision values and prevent distinct path values from collapsing into the same level, ensuring the system remains sensitive to variations in the total cost.

There is another key element that must be clarified, the operator used in Phase Estimation to introduce the cost values as phases. We will stick to~\cite{srinivasan2018efficient}, to that end we can construct a matrix $A$ such that $A_{ij}=e^{-2\pi i\phi_{ij}}$ where $\phi_{ij}$ represents the cost of traveling from city $i$ to city $j$ after normalization. The final operator ($U$) is created by splitting the information in $A$ into smaller operators. Each of these operators will act on each of the $m$ groups of qubits representing each city. The new operators ($Q^j$) are constructed as diagonal matrices, of dimension $2^{\lceil \log_2(m) \rceil}$, whose diagonal values can be subtracted from the $j$-th column of matrix $A$. Note that the values on the $Q^j$ matrix represent the cost of traveling from all different cities to the one selected ($j$). Under this definition, we have $m$ diagonal operators $Q^j$ such that:
\begin{equation}
    Q^j_{ii}=A_{ij}.
\end{equation}
The operator $U$ is expressed in terms of different and independent operations on the distinct $m$ groups of qubits as:
\begin{equation}
   U= \bigotimes_{j=0}^{m-1}   Q^j.
   \label{uoperator}
\end{equation}
For the sake of clarity, we can recover the four cities example. It is interesting to have a look at one of the operators $Q^j$, as is the case of:
\begin{equation}
    Q^0=\ketbra{00}{00}+e^{-2\pi i\phi_{10}}\ketbra{01}{01}+e^{-2\pi i\phi_{20}}\ketbra{10}{10}+e^{-2\pi i\phi_{30}}\ketbra{11}{11}.
\end{equation}
We can also show the application of the $U$ operator. To this end, we can reuse the route $\ket{10110100}$  
\begin{equation}
    U\ket{10110100}=Q^0\otimes Q^1 \otimes Q^2\otimes Q^3\ket{10110100}=e^{-2\pi i(\phi_{20}+\phi_{31}+\phi_{12}+\phi_{03})}\ket{10110100}.
\end{equation}
It is clearly seen that the total route cost appears as the phase in the eigenvalue. We must make a final remark in the construction of $U$, if we revisit Eq.~\ref{phase est} we are applying controlled operations of $U$. This means that we need the definition of $U$ to hold true for $C\text{-}U$, we found that trivial given $C\text{-}(A\otimes I) \cdot C\text{-}(I\otimes B)= C\text{-}(A\otimes B)$, which implies:
\begin{equation}
    C\text{-}U= \prod_{j=0}^{m-1} C\text{-}(I^{\otimes j }\otimes  Q^j \otimes I^{\otimes (m-j-1) }) .
\end{equation}
Therefore, in practice, it is sufficient to apply the different controlled gates $Q^j$ to the corresponding groups of qubits. Furthermore, this diagonal structure allows the Phase Estimation stage to be executed more efficiently. While finding an efficient implementation for the controlled powers of the unitary ($C\text{-}U^{2^k}$) is often non-trivial, frequently requiring a massive stack of consecutive gates that prohibitively increases circuit depth, this framework avoids this structural circuit growth; because each $Q^j$ is diagonal, raising the operator to successive powers of two does not lead to structural circuit growth. Instead, it merely requires scaling the phase angles encoded within the diagonal elements.

We now have all the ingredients to start formulating our algorithm. The first subroutine will be called initiator ($T$), and it consists of the quantum circuit needed to create a uniform superposition of the states $\ket{C}\otimes \ket{R}$. As has been previously stated, this is made through Phase Estimation. The first step of $T$ consists of creating the superposition along all possible routes on the $\ket{R}$ register $T_1\ket{0}\ket{0}=\frac{1}{\sqrt{(m-1)!}}\sum_r \ket{0}\ket{r}$. In this work, as a way to create such states, we first generate a uniform superposition among the first $(m-1)!$ states via~\cite{shukla2024efficient} and subsequently, using the method in~\cite{ramacciotti2024simple}, the amplitudes of these first states are permuted with the states encoding the routes; although a more efficient implementation is fundamental to the search for quantum advantage. The next step is to perform Phase Estimation using the $U$ operator from Eq.~\ref{uoperator}. Just as a reminder, it consists of applying Hadamard gates over all the $\ket{C}$ register, then making successive controlled operations of the different $U^{2^k}$ powers of $U$, and finally performing the QFT$^{-1}$ over the $\ket{C}$ register. After Phase Estimation we have
\begin{equation}
T_2\frac{1}{\sqrt{(m-1)!}}\sum_r \ket{0}\ket{r}=\frac{1}{\sqrt{(m-1)!}}\sum_r \ket{c_r}\ket{r}.
\label{estaodosmeh}
\end{equation}
Note that, in reality, the states will not be this "perfect", as not every cost is divisible by $2^d$ in general. This means that, in actual scenarios, each route will result in a distinct set of costs and variations in their respective amplitudes, falling around the actual cost. However, as will be shown later, this phenomenon is of no consequence to the dynamics of the algorithm. For simplicity we have decided to stick with the definition $T=T_2T_1$ as:
\begin{equation}
    T\ket{0}\ket{0}=\frac{1}{\sqrt{(m-1)!}}\sum_r \ket{c_r}\ket{r}.
    \label{tapplication}
\end{equation}
Once we have an operator $T$ that allows us to create the states $\ket{C}\otimes \ket{R}$ we can use it as the initiator in the Grover-Long algorithm. This means that we can perform a Grover-long algorithm where $Q=-TI_0T^{-1}I_\mu$. As we have already mentioned, the Grover-Long algorithm will be used as a minimum finder, more specifically we will follow the work of~\cite{liu_optimized_2021}. The idea is to select a threshold for the cost value, and then mark and amplify the amplitude of the  states with a cost lower than that threshold value. After the Grover-Long algorithm is applied, a new cost value is measured and selected as a new threshold. This process is repeated until we find the minimum. The minimum is identified by measuring the same state repeatedly $k$ times. Note that at any intermediate threshold step, if $n$ states satisfy the current cost condition, the probability of measuring any specific state among them is equal to $1/n$. Therefore, the probability of measuring the same non-unique state $k$ consecutive times is $1/n^k$. Since the exact number of states $n$ remaining below the threshold is unknown, we apply the bound over all possible valid subspace sizes $n \in [2, (m-1)!]$ to account for the worst-case scenario. Consequently, the probability $P$ that the measured state after $k$ identical consecutive outcomes is indeed the minimum state is bounded by 
\begin{equation}
    P \ge  1- \sum_{n=2}^{(m-1)!}\frac{1}{n^{k}}.
\end{equation}
Where the subtracted summation represents the probability of a false positive, accounting for the probability of consecutively measuring the same state $k$ times when, in reality, it is not unique. There is one last detail on the application of the Grover-Long algorithm that needs to be addressed. As proven in Section~\ref{seccion pe}, in order to perform the Grover-Long algorithm it is imperative to have knowledge about the number of solutions $\Big(\sin(\beta)=\sqrt{\frac{N'}{N}} \Big)$. However, in general this information is unknown, as we do not know how many routes have lower cost value than the current threshold. Instead of $\frac{N'}{N}$ we will make an estimation using the total number of states accessible in the register $\ket{C}$ and the cost threshold ($c_t$). Thus, we have $\sin(\beta)=\sqrt{\frac{c_t}{2^d}}$. It is worth noting, however, that utilizing this approximation prevents us from strictly guaranteeing the theoretical $100\%$ success rate, as the deviation from the true ratio can lead to a slight over or under rotation in the Amplitude Amplification process. We can also remember now the issue with Eq.~\ref{estaodosmeh}, where for each route we have slightly different cost values. However, given we take $d$ big enough, during the process of minimization it is irrelevant if the cost measured does not match exactly with the cost value, what is important is that the routes with lower cost values will still be marked and the iterative process persists without any problem.

Finally, we can give a brief description of the computational complexity of the algorithm. Given the uniform probability of obtaining each of the solutions with cost value less than the threshold, the expected number of iterations before we reach the minimum is $\log_2((m-1)!)$. Using Stirling's approximation, and restricting ourselves to the higher factor, it can be seen as $\mathcal{O}(m\log_2(m))$. In each iteration we are repeating the Grover-Long operator $n\ge \frac{\frac{\pi}{2}-\beta}{2\beta}$ times, which in the worst case is defined by $n\ge \frac{\frac{\pi}{2}-\arcsin(\sqrt{\epsilon})}{2\arcsin(\sqrt{\epsilon})}$. In the limit where $\epsilon \ll 1$, we have $n_{max} \approx\sqrt{\frac{1}{\epsilon}}$. finally, we must find the element that gives rise to higher complexity on the Grover-Long operator. In this work, the $T_1$ operator yields a high scaling of $\mathcal{O} \big ((m-1)! \log_2[(m-1)!]\big)$, based on~\cite{shukla2024efficient} and~\cite{ramacciotti2024simple}. However, we do not consider this operator in our final evaluation for two main reasons. First, in our current approach, $T_1$ is treated as a generic sparse vector, leaving its highly defined mathematical structure unexploited. Second, its complexity depends solely on the number of cities $m$; thus, any efficient implementation developed for a specific city count can be universally applied to all instances of the same size. Consequently, we anticipate that future algorithmic advancements will leverage these properties to optimize $T_1$, allowing us to treat Phase Estimation as the primary driver of complexity and confine our evaluation solely to this process. In order to analyze QPE complexity we must take into account that each one of our operators $U^{2^x}$ is constructed as $m$ diagonal operators. As we are applying $m \cdot d$ diagonal operators, and assuming that the computational cost of applying each $(Q^j)^{2^x}$ is constant $\mathcal{O}(1)$, we can take $\mathcal{O}(m \log_2(1/\epsilon))$ as our Phase Estimation induced complexity. We are now in position to construct our expected complexity as $\mathcal{O}(\frac{m^2\log_2(m)\log_2(1/\epsilon)}{\sqrt{\epsilon}})$. Since $d$ scales as $\mathcal{O}(\log_2(m))$, which in turn implies that $\epsilon$ scales as $\mathcal{O}(1/m)$, the overall complexity simplifies exclusively in terms of $m$ to $\mathcal{O}(m^{5/2}\log^2(m))$. While this evaluation relies on an expectation value and omits the explicit overhead of the $T_1$ preparation stage, it nonetheless provides valuable insights into the scaling behavior of the algorithm.

%% file: practicalcase.tex
\begin{figure}[h]
    \centering
    \includegraphics[width=0.5\linewidth]{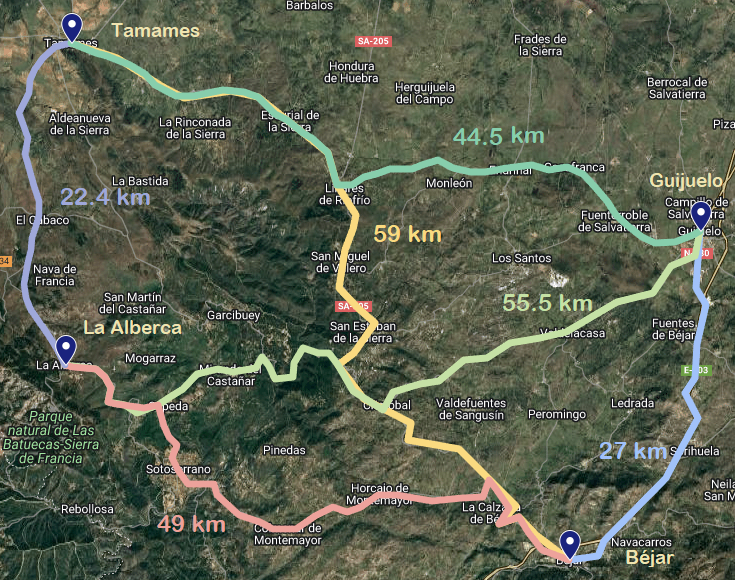}
    \caption{Map presenting the four towns of Salamanca we are using as toy model and the shortest possible path between them.}
    \label{mapa}
\end{figure}

In this section, we present an example of the application of the Traveling Salesman Problem iterative solver. 
First of all, it is important to emphasize that the quantum algorithms were programmed using Qiskit~\cite{Qiskit} and simulated rather than executed on actual quantum hardware, utilizing the Qiskit Aer ideal simulator for all executions, sampling 1 shot per implementation step and using a stopping parameter of $k = 4$ identical consecutive evaluations. The case study will utilize four towns in the province of Salamanca (Spain) as a toy model. Although the test model is highly simplified due to the high computational cost associated with quantum simulations, it provides valuable insight into the fundamental workings of the algorithm, laying the core conceptual and operational foundations of the proposed framework. The towns and individual path costs are represented in Figure~\ref{mapa} and Table~\ref{kilometers} and the routes cost and encoding can be seen in Table~\ref{routescosts}. Note that, as we are in a symmetric TSP problem, the direction in which the route is traveled is irrelevant. Therefore, when implementing the algorithm, measuring either of these two routes will be considered the same state.  To normalize the cost values, we apply a factor of $1/(4 \times 59)$, determined by the number of cities (4) and the maximum inter-city route cost (59). Regarding the $\ket{C}$ register, an accuracy parameter of $d = 6$ is used, resulting in a total resource requirement of 14 qubits.

\begin{table}[ht]
            \caption{ Information about the encoding used for towns and routes, and values for the costs of paths and routes in the symmetric Salamanca toy model.}
    \label{table:cost info}
    \begin{subtable}{\textwidth}
        \centering
            \subcaption{Distances in kilometers by the shortest path between the four different towns.}
            \label{kilometers}
        \begin{tabular}{|c|r|r|r|r|}
            \hline
                       & Guijuelo (00) & Tamames (01) & La Alberca (10) & Béjar (11) \\ \hline
            Guijuelo (00) & 0.0        & 44.5    & 55.5       & 27.0    \\ \hline
            Tamames  (01)  & 44.5     & 0.0      & 22.4       & 59.0    \\ \hline
             La Alberca (10) & 55.5   & 22.4    & 0.0          & 49.0    \\ \hline
            Béjar (11)    & 27.0       & 59.0      & 49.0         & 0.0     \\ \hline
        \end{tabular}

    \end{subtable}

    \vspace{\baselineskip}
        
    \begin{subtable}{\textwidth}
        \subcaption{Encoding and cost in kilometers for the six different routes.}
        \label{routescosts}
    
        \centering
        \begin{tabular}{|c|c|r|}
            \hline
            Route & Encoding     & Cost (km) \\ \hline
           $ 00 \to 01 \to 10 \to 11\to 00$& 11000110 & 142.9     \\ \hline
           $ 00 \to 01 \to 11 \to 10\to 00$& 10001101 & 208.0       \\ \hline
           $ 00 \to 10 \to 01 \to 11\to 00$& 11100001 & 163.9     \\ \hline
           $ 00 \to 10 \to 11 \to 01 \to 00$& 01110010 & 208.0       \\ \hline
           $ 00 \to 11 \to 01 \to 10 \to 00$& 10110100 & 163.9     \\ \hline
           $ 00 \to 11 \to 10 \to 01 \to 00$& 01101100 & 142.9     \\ \hline
        \end{tabular}

    \end{subtable}
\end{table}

The bitstrings $11000110$ and $01101100$ represent both directions of Guijuelo $\leftrightarrows $ Tamames $\leftrightarrows$ La Alberca $\leftrightarrows$ Béjar $\leftrightarrows$ Guijuelo, which are indeed the shortest possible routes.  To enhance the understanding of the algorithm, we will now analyze various executions to uncover the underlying logic. The random nature of quantum computing renders distinct paths for the algorithm to reach the minimum, as evidenced by the two executions depicted in Table~\ref{examplesexecutions}. It is noteworthy that the initial route and cost values are selected randomly through direct measurement of the state in Eq.~\ref{tapplication}, thereby yielding disparate first thresholds. The two examples demonstrate how the number of iterations required to identify the minimum can vary depending on the state selected in each step. We can also revisit the issue of the cost not matching its exact value in Eq.~\ref{estaodosmeh}. If we compare the values on Table~\ref{routescosts} and Table~\ref{examplesexecutions} the cost values are different. In fact, in Table~\ref{examplesexecutions} there are different cost values for the same route. However, it is worth noting that during intermediate steps, any slight cost fluctuations do not affect the algorithm’s performance, as the iterative process simply continues to narrow down the threshold. The sole critical point occurs at the final selected cost threshold, where route mixing may manifest. Nevertheless, since the resolution scales with the Phase Estimation precision parameterized by $\epsilon$, any overlapping routes will generally differ by an amount comparable to this error scale. Consequently, given our escalation of $\epsilon$ based on the bit-growth effect, any mixed routes remain closely aligned in cost. This behavior is clearly illustrated by this symmetric case discussed; since both directions of the same tour share the exact same cost, they naturally overlap within this probabilistic resolution.

\begin{table}[ht]
\centering
\caption{Two different examples of executions selected to show different scenarios when the TSP iterative solver has been tested on the symmetric toy model.}
\label{examplesexecutions}
\begin{tabular}{c|cc|cc|}
          & \multicolumn{2}{c|}{Example 1} & \multicolumn{2}{c|}{Example 2} \\ \hline
Iteration & Cost          & Route          & Cost          & Route          \\ \hline
1         & 165.95        & 11100001       & 206.50        & 10001101       \\
2         & 143.81        & 11000110       & 165.95        & 10110100       \\
3         & 143.81        & 01101100       & 162.24        & 10110100       \\
4         & 140.12        & 01101100       & 143.81        & 01101100       \\
5         & 140.12        & 11000110       & 143.81        & 11000110       \\
6         &               &                & 143.81        & 01101100       \\
7         &               &                & 143.81        & 01101100      
\end{tabular}
\end{table}

Now, we can extend our analysis to the asymmetric Traveling Salesman Problem by considering the same four towns in the province of Salamanca, Spain. This time, we introduce asymmetry in the travel costs, incorporating directionality in the route. The travel cost from one town to another is no longer necessarily the same in both directions. By addressing this more complex and nuanced version of the TSP, we enhance the clarity of our analysis, as we now explore a scenario with added intricacies compared to the symmetric TSP case. To introduce this asymmetry, we account for the altitude differences between the cities. The travel cost is then adjusted based on the elevation changes, providing a more realistic representation of the travel expenses. The new cost can be obtained in terms of the symmetric cost as:
\begin{equation}
    c' = \frac{c}{1-\Delta y/2}.
\end{equation}
Where  $\Delta y$   represents the altitude difference between the two towns. This formulation is chosen simply as a practical model to capture directional effects: a positive value increases the cost for uphill travel $c' > c$, while a negative value reduces it for downhill paths $c' < c$. To maintain consistency with the previous case, we continue using the factor of $1/(4 \times 59)$. Even though now there are paths with a higher cost, as this is a small instance of the TSP, we can easily confirm that everything remains properly normalized. Furthermore, as the route costs now exhibit smaller differences and we aim for the algorithm to distinguish these subtler variations,  we set the number of qubits in the  $\ket{C}$ register to $ d = 7$, bringing the total qubit count to 15 qubits. In this modified problem setup, the different path and route costs are provided in Table~\ref{table:cost info asym}. Notably, there is now a single route with the minimum cost, represented by the bitstring  11000110, which corresponds to the route: Guijuelo $ \rightarrow $ Tamames $ \rightarrow $ La Alberca $ \rightarrow $ Béjar $ \rightarrow $ Guijuelo. As in this new context the algorithm can take many different paths to find the solution, we have decided to examine three example runs of the algorithm, presented in Table~\ref{examplesexecution asym}. A similar outcome can be observed as in the symmetric case, where the algorithm progressively narrows the solution space, ultimately identifying the route with the lowest cost. Furthermore, by enhancing the precision through the addition of a qubit to the $\ket{C}$ register, the fluctuations in costs, resulting from the aforementioned issue in Eq.~\ref{estaodosmeh}, are less pronounced in these new runs. The empirical results detailed in Table~\ref{examplesexecutions} and Table~\ref{examplesexecution asym} validate the performance of the iterative solver across both symmetric and asymmetric scenarios. These practical use cases effectively illustrate the inner dynamics and operational workflow of the proposed quantum-classical iterative framework.

\begin{table}[ht]
            \caption{ Information about the encoding used for towns and routes, and values for the costs of paths and routes in the non-symmetric Salamanca toy model.}
    \label{table:cost info asym}
    \begin{subtable}{\textwidth}
        \centering
            \subcaption{Costs after taking into account the altitude for the paths between the four different towns.}
            \label{kilometers asym}
\begin{tabular}{|c|r|r|r|r|}
    \hline
                       & Guijuelo (00) & Tamames (01) & La Alberca (10) & Béjar (11) \\ \hline
    Guijuelo (00)      & 0.00          & 41.61        & 58.39            & 27.84      \\ \hline
    Tamames (01)        & 47.82        & 0.00         & 25.43          & 65.55     \\ \hline
    La Alberca (10)     & 52.88         & 20.01       & 0.00             & 48.08      \\ \hline
    Béjar (11)          & 26.20        & 53.64        & 49.94          & 0.00       \\ \hline
\end{tabular}

    \end{subtable}

    \vspace{\baselineskip}
        
    \begin{subtable}{\textwidth}
        \subcaption{Encoding and cost  for the six different routes.}
        \label{routescosts asym}
    
        \centering
        \begin{tabular}{|c|c|r|}
            \hline
            Route & Encoding     & Weighted Cost \\ \hline
           $ 00 \to 01 \to 10 \to 11\to 00$& 11000110 & 141.32     \\ \hline
           $ 00 \to 01 \to 11 \to 10\to 00$& 10001101 & 209.97       \\ \hline
           $ 00 \to 10 \to 01 \to 11\to 00$& 11100001 & 170.15     \\ \hline
           $ 00 \to 10 \to 11 \to 01 \to 00$& 01110010 & 207.93       \\ \hline
           $ 00 \to 11 \to 01 \to 10 \to 00$& 10110100 & 159.79     \\ \hline
           $ 00 \to 11 \to 10 \to 01 \to 00$& 01101100 & 145.61     \\ \hline
        \end{tabular}

    \end{subtable}
\end{table}
\begin{table}[h]
\centering
\caption{Example executions when the TSP iterative solver has been tested on the non-symmetric toy model.}
\label{examplesexecution asym}
\begin{tabular}{c|cc|cc|cc|}
          & \multicolumn{2}{c|}{Example 1} & \multicolumn{2}{c|}{Example 2} &\multicolumn{2}{c|}{Example 3}\\ \hline
Iteration & Cost          & Route          & Cost          & Route         &Cost          & Route \\ \hline
1         & 206.50        & 01110010       & 208.34      & 01110010     &210.18 &10001101  \\
2         & 206.50        & 01110010       & 169.63        & 11100001    &208.34&01110010   \\
3         & 145.66        & 01101100      & 141.97        & 11000110     &156.71&10110100  \\
4         & 141.97        & 11000110       & 141.97        &11000110    &141.97        & 11000110   \\
5         & 141.97        & 11000110       & 140.13        & 11000110     &141.97        & 11000110 \\
6         & 141.97         &  11000110              & 140.13        & 11000110  &141.97        & 11000110    \\
7         & 141.97              &  11000110              &         &     &141.97        & 11000110
\end{tabular}
\end{table}

%% file: conclusion.tex
In this study, we introduced a hybrid quantum-classical iterative framework designed to tackle the Traveling Salesman Problem  by capitalizing on the synergy between Quantum Phase Estimation and Quantum Amplitude Amplification via the Grover-Long algorithm. By encoding normalized tour distances as quantum phases through a tailored diagonal unitary operator, this architecture successfully circumvents the structural circuit depth explosion typically associated with controlled unitary powers. The outer classical loop drives the algorithmic optimization, systematically narrowing the solution space by adapting the cost threshold based on successive quantum measurements. The core operational principles of this approach were validated using both symmetric and asymmetric four-city toy models simulated on the Qiskit Aer framework. The empirical results confirm that the algorithm successfully converges toward the global minimum tour in both configurations. While these preliminary results demonstrate a mathematically sound pathway toward quantum optimization, transitioning from a proof-of-concept simulation to a scalable framework requires addressing key algorithmic and physical bottlenecks in subsequent research.\\
A primary avenue for future work involves the structural optimization of the state preparation operator $T_1$. In its current formulation, this operator introduces a severe factorial bottleneck, primarily because it treats the target states as a generic sparse vector. However, $T_1$ possesses a highly defined, rigid mathematical structure that remains unexploited in this study. Future efforts should explore advanced state preparation subroutines that explicitly leverage these geometric symmetries to compress the initialization circuit and significantly reduce its overhead.  Additionally, evaluating the robustness of this framework under quantum noise and physical deployments represents a critical next step. While the ideal simulations verified the validity of the algorithm, executing this framework on real hardware is expected to introduce severe physical challenges. Finally, scaling the current framework to larger problem sizes ($m > 4$) will be essential to evaluate the real-world performance of the probabilistic threshold-based minimum-finding loop under massive search spaces.

%% file: matematicasgroverlong.tex
In this supplementary material, a description of the $n$-th application of Grover-Long operator is provided. While this section does not introduce novel theoretical results, it is included here for the sake of completeness and to ensure the manuscript remains entirely self-contained. To this end, we will follow the work of~\cite{toyama2008multiphase,toyama_quantum_2013}. Our starting point will be the $Q$ operator we had in Section~\ref{amplitudamplification sec}:
\begin{equation}
    Q=\begin{bmatrix} -e^{i \phi} (1+(e^{i \phi}-1)\sin^2{\beta}) &-(e^{i \phi}-1) \sin{\beta} \cos{\beta}  \\-e^{i \phi} (e^{i \phi}-1) \sin{\beta} \cos{\beta}  & -e^{i \phi} + (e^{i \phi}-1)\sin^2{\beta}  \end{bmatrix}.
\end{equation}
In order to align with~\cite{toyama2008multiphase} we will add a global phase $-e^{-i\phi}$.
\begin{equation}
    G=-e^{-i\phi} Q= \begin{bmatrix} (1+(e^{i \phi}-1)\sin^2{\beta}) &-(e^{-i \phi}-1) \sin{\beta} \cos{\beta}  \\ 
    (e^{i \phi}-1) \sin{\beta} \cos{\beta}  & 1 + (e^{-i \phi}-1)\sin^2{\beta}  \end{bmatrix}.
\end{equation}
We now determine the characteristic polynomial of $G$ given by $ \det (\lambda I-G)=0$.
\begin{equation}
\begin{split}
   \lambda^2-2\lambda -(e^{i \phi}-1)\sin^2(\beta)\lambda-(e^{-i \phi}-1)\sin^2(\beta)\lambda+(e^{i \phi}-1)\sin^2(\beta)+(e^{-i \phi}-1)\sin^2(\beta)+& \\ +1 +(e^{-i \phi}-1)(e^{i \phi}-1)\sin^4(\beta)+(e^{-i \phi}-1)(e^{i \phi}-1)\sin^2(\beta)\cos^2(\beta)=0&.
    \end{split}
\end{equation}
After some algebra, this can be expressed as:
\begin{equation}
    \lambda^2+2\lambda(-1+(1-\cos\phi)\sin^2(\beta))+1=0.
    \label{eq caracteristica}
\end{equation}
Solving for $\lambda$ yields:
\begin{equation}
    \lambda=1-(1-\cos\phi)\sin^2(\beta)\pm \sqrt{(-1+(1-\cos\phi)\sin^2(\beta))^2-1}.
\end{equation}
Making the transformation $x=(1-\cos\phi)\sin^2(\beta)$ we have:
\begin{equation}
    \lambda=1-x\pm i\sqrt{x(2-x)}.
\end{equation}
With this definition it can be seen that $\lambda=e^{\pm i \theta}$, where:
\begin{equation}
    \theta=\arctan \Bigg (\frac{\sqrt{x(2-x)}}{1-x} \Bigg ). 
\end{equation}
Equation~\ref{eq caracteristica} can thus be rewritten as:
\begin{equation}
    \lambda^2-2\lambda \cos(\theta)+1=0.
\end{equation}
Making use of the Cayley-Hamilton theorem, we know that this equation is also fulfilled by the operator $G$,
\begin{equation}
    G^2=2G \cos(\theta)- I.
\end{equation}
As shown in~\cite{sprung_scattering_1993} this defines a recurrence relation for $G^n$:
\begin{equation}
    G^n=\frac{1}{\sin \theta} \bigg (G \sin (n\theta)-\sin((n-1)\theta)I \bigg ).
\end{equation}
Returning to our initial operator, we obtain:
\begin{equation}
    Q^n=\frac{(-1)^ne^{i\phi n}}{\sin \theta} \bigg (-e^{-i\phi}Q \sin (n\theta)-\sin((n-1)\theta) I \bigg ).
\end{equation}
This allows the amplitudes of the $\ket{\mu}$ and $\ket{\tau}$ states to be recovered as:
\begin{equation}
    \begin{bmatrix} A_\mu \\  A_\tau \end{bmatrix}  = Q^n \begin{bmatrix} \sin \beta \\  \cos \beta \end{bmatrix}.
\end{equation}
After algebraic manipulation, we determine the amplitude $A_\tau$.
\begin{equation}
    A_\tau=\frac{(-1)^{n}e^{i\phi n}\cos \beta}{\sin \theta} \bigg( (1-2x)\sin( n\theta) -\sin((n-1)\theta) \bigg ).
\end{equation}
We can use the subtraction formula for sine to rewrite it as:
\begin{equation}
   A_\tau=\frac{(-1)^{n}e^{i\phi n}\cos \beta}{\sqrt{x(2-x)}} \bigg(\sin( n\theta)(1-2x-\cos(\theta))+\cos(n\theta)\sin(\theta) \bigg).
\end{equation}
If we use a new variable such as $\tan \alpha= \sqrt{\frac{x}{2-x}}$ with $\sin \alpha=\sqrt{\frac{x}{2}}$,
\begin{equation}
    A_\tau=\frac{(-1)^{n}e^{i\phi n}\sqrt{2}\cos \beta}{\sqrt{2-x}} \cos(n \theta+\alpha).
\end{equation}
Note that $\sin(2 \alpha)=\sin(\theta)$, which implies $\alpha=\theta/2+ k\pi \ \forall k \in \mathbb{Z}$, thus we can rewrite $A_\tau$,
\begin{equation}
      A_\tau=\frac{(-1)^{n}e^{i\phi n}\cos \beta}{\cos(\theta/2+k\pi)} \cos((n+1/2) \theta+k\pi).
\end{equation}
Since the success probability is $P=1-\abs{A_\tau}^2$, achieving $P=1$ requires:
\begin{equation}
    \theta=\frac{\pi}{2n+1}.
\end{equation}
Using $\cos(\theta)=1-((1-\cos\phi)\sin^2(\beta))$ we have:
\begin{equation}
    \phi=\arccos\bigg ( 1 -\frac{1-\cos(\frac{\pi}{2n+1})}{\sin^2(\beta)} \bigg).
\end{equation}
We have found a relation between the angle $\phi$ and the number of iterations $n$ that yields a success probability of $P=1$. This can be rewritten to align with the original paper~\cite{long2001grover}, by using $\sin(\phi/2) = \sqrt{\frac{1 - \cos(\phi)}{2}}$ twice.
\begin{equation}
    \phi= 2 \arcsin\bigg(\frac{\sin(\frac{\pi}{4n+2})}{\sin(\beta)} \bigg).
\end{equation}
Note that for the $\arcsin$ function to be well-defined, its argument must lie within the interval $[-1,1]$, this means that taking $\beta \in [0,\frac{\pi}{2}]$ the condition $\beta \ge \frac{\pi}{4n+2}$ must be fulfilled. This leads to a minimum number of iterations:
\begin{equation}
    n \ge \frac{\frac{\pi}{2}-\beta}{2\beta}.
\end{equation}

%% file: BiblioTSP.bib
@article{coppersmith2002approximate,
  author  = {Coppersmith, Don},
  title   = {An approximate {Fourier} transform useful in quantum factoring},
  journal = {arXiv preprint quant-ph/0201067},
  year    = {2002}
}

@book{garey2002computers,
  author    = {Garey, Michael R. and Johnson, David S.},
  title     = {Computers and Intractability: {A} Guide to the Theory of {NP}-Completeness},
  publisher = {W. H. Freeman},
  address   = {New York},
  year      = {2002}
}

@article{gharehchopogh2023improved,
  title={An improved farmland fertility algorithm with hyper-heuristic approach for solving travelling salesman problem},
  author={Gharehchopogh, Farhad and Abdollahzadeh, Benyamin and Arasteh, Bahman},
  journal={Computer Modeling in Engineering \& Sciences},
  volume={135},
  number={3},
  pages={1981},
  year={2023},
  publisher={Tech Science Press}
}

@misc{Qiskit,
    author = {{Qiskit contributors}},
    title = {Qiskit: An Open-source Framework for Quantum Computing},
    year = {2019},
    doi = {10.5281/zenodo.2562110},
    url = {https://github.com/Qiskit/qiskit}
}

@article{oppenheim1972effects,
  author  = {Oppenheim, Alan V. and Weinstein, Clifford J.},
  title   = {Effects of finite register length in digital filtering and the fast {Fourier} transform},
  journal = {Proceedings of the {IEEE}},
  volume  = {60},
  number  = {8},
  pages   = {957--976},
  year    = {1972}
}

@article{zhu2022realizable,
  author  = {Zhu, Jieao and Gao, Yihuai and Wang, Hansen and Li, Tiefu and Wu, Hao},
  title   = {A realizable gas-based quantum algorithm for traveling salesman problem},
  journal = {arXiv preprint arXiv:2212.02735},
  year    = {2022}
}

@article{shukla2024efficient,
  author  = {Shukla, Alok and Vedula, Prakash},
  title   = {An efficient quantum algorithm for preparation of uniform quantum superposition states},
  journal = {Quantum Information Processing},
  volume  = {23},
  number  = {2},
  pages   = {38},
  year    = {2024}
}

@article{ramacciotti2024simple,
  title={Simple quantum algorithm to efficiently prepare sparse states},
  author={Ramacciotti, Debora and Lefterovici, Andreea I and Rotundo, Antonio F},
  journal={Physical Review A},
  volume={110},
  number={3},
  pages={032609},
  year={2024},
  publisher={APS}
}

@article{gharehchopogh2022efficient,
  author  = {Gharehchopogh, Farhad Soleimanian and Abdollahzadeh, Benyamin},
  title   = {An efficient {Harris} hawk optimization algorithm for solving the travelling salesman problem},
  journal = {Cluster Computing},
  volume  = {25},
  number  = {3},
  pages   = {1981--2005},
  year    = {2022}
}

@article{lucas2014ising,
  author  = {Lucas, Andrew},
  title   = {{Ising} formulations of many {NP} problems},
  journal = {Frontiers in Physics},
  volume  = {2},
  pages   = {74887},
  year    = {2014}
}

@article{farhi2014quantum,
  title={A quantum approximate optimization algorithm},
  author={Farhi, Edward and Goldstone, Jeffrey and Gutmann, Sam},
  journal={arXiv preprint arXiv:1411.4028},
  year={2014}
}

@article{ruan2020quantum,
  title={The quantum approximate algorithm for solving traveling salesman problem},
  author={Ruan, Yue and Marsh, Samuel and Xue, Xilin and Liu, Zhihao and Wang, Jingbo},
  journal={Computers, Materials, \& Continua},
  volume={63},
  number={3},
  pages={1237},
  year={2020},
  publisher={Tech Science Press}
}

@article{abdollahzadeh2021discrete,
  author  = {Abdollahzadeh, Benyamin and Soleimanian Gharehchopogh, Farhad and Barshandeh, Saeid},
  title   = {Discrete farmland fertility optimization algorithm with metropolis acceptance criterion for traveling salesman problems},
  journal = {International Journal of Intelligent Systems},
  volume  = {36},
  number  = {3},
  pages   = {1270--1303},
  year    = {2021}
}

@incollection{applegate2011traveling,
  author    = {Applegate, David L. and Bixby, Robert E. and Chv{\'a}tal, Va{\v{s}}ek and Cook, William J.},
  title     = {The traveling salesman problem: a computational study},
  booktitle = {The Traveling Salesman Problem},
  publisher = {Princeton University Press},
  year      = {2011}
}

@article{dorigo2002ant,
  author  = {Dorigo, Marco and Gambardella, Luca Maria},
  title   = {Ant colony system: a cooperative learning approach to the traveling salesman problem},
  journal = {{IEEE} Transactions on Evolutionary Computation},
  volume  = {1},
  number  = {1},
  pages   = {53--66},
  year    = {1997}
}

@inproceedings{ulder1990genetic,
  author    = {Ulder, Nico L. J. and Aarts, Emile H. L. and Bandelt, Hans-J{\"u}rgen and Van Laarhoven, Peter J. M. and Pesch, Erwin},
  title     = {Genetic local search algorithms for the traveling salesman problem},
  booktitle = {International Conference on Parallel Problem Solving from Nature},
  pages     = {109--116},
  publisher = {Springer},
  year      = {1990}
}

@inproceedings{applegate1998solution,
  author    = {Applegate, David and Bixby, Robert and Chv{\'a}tal, Va{\v{s}}ek and Cook, William},
  title     = {On the solution of traveling salesman problems},
  booktitle = {Proceedings of the International Congress of Mathematicians 1998},
  pages     = {645--656},
  publisher = {European Mathematical Society-EMS-Publishing House GmbH},
  year      = {1998}
}

@article{brassard2002quantum,
  title={Quantum amplitude amplification and estimation},
  author={Brassard, Gilles and Hoyer, Peter and Mosca, Michele and Tapp, Alain},
  journal={Contemporary Mathematics},
  volume={305},
  pages={53--74},
  year={2002},
  publisher={Providence, RI: American Mathematical Society}
}

@article{warren2013adapting,
  author  = {Warren, Richard H.},
  title   = {Adapting the traveling salesman problem to an adiabatic quantum computer},
  journal = {Quantum Information Processing},
  volume  = {12},
  number  = {4},
  pages   = {1781--1785},
  year    = {2013}
}

@article{goswami2024solving,
  title={Solving the travelling salesman problem using a single qubit},
  author={Goswami, Kapil and Veereshi, Gagan Anekonda and Schmelcher, Peter and Mukherjee, Rick},
  journal={arXiv preprint arXiv:2407.17207},
  year={2024}
}

@techreport{balas1983branch,
  author      = {Balas, Egon and Toth, Paolo},
  title       = {Branch and bound methods for the traveling salesman problem},
  institution = {Carnegie-Mellon University Management Sciences Research Group},
  type        = {Management Sciences Research Report},
  number      = {MSRR-488},
  year        = {1983}
}

@article{martovnak2004quantum,
  title={Quantum annealing of the traveling-salesman problem},
  author={Marto{\v{n}}{\'a}k, Roman and Santoro, Giuseppe E and Tosatti, Erio},
  journal={Physical Review E—Statistical, Nonlinear, and Soft Matter Physics},
  volume={70},
  number={5},
  pages={057701},
  year={2004},
  publisher={APS}
}

@article{bellman_dynamic_1962,
	title = {Dynamic {Programming} {Treatment} of the {Travelling} {Salesman} {Problem}},
	volume = {9},
	issn = {0004-5411},
	url = {https://dl.acm.org/doi/10.1145/321105.321111},
	doi = {10.1145/321105.321111},
	number = {1},
	urldate = {2026-01-12},
	journal = {J. ACM},
	author = {Bellman, Richard},
	month = jan,
	year = {1962},
	pages = {61--63},
}

@article{lin1973effective,
  author  = {Lin, Shen and Kernighan, Brian W.},
  title   = {An effective heuristic algorithm for the traveling-salesman problem},
  journal = {Operations Research},
  volume  = {21},
  number  = {2},
  pages   = {498--516},
  year    = {1973}
}

@inproceedings{shor_algorithms_1994,
	title = {Algorithms for quantum computation: discrete logarithms and factoring},
	shorttitle = {Algorithms for quantum computation},
	url = {https://ieeexplore.ieee.org/document/365700},
	doi = {10.1109/SFCS.1994.365700},
	urldate = {2024-10-07},
	booktitle = {Proceedings 35th {Annual} {Symposium} on {Foundations} of {Computer} {Science}},
	author = {Shor, P.W.},
	month = nov,
	year = {1994},
	pages = {124--134},
}

@article{kitaev1995quantum,
  title={Quantum measurements and the Abelian stabilizer problem},
  author={Kitaev, A Yu},
  journal={arXiv preprint quant-ph/9511026},
  year={1995}
}

@article{srinivasan2018efficient,
  author  = {Srinivasan, Karthik and Satyajit, Saipriya and Behera, Bikash K. and Panigrahi, Prasanta K.},
  title   = {Efficient quantum algorithm for solving travelling salesman problem: {An} {IBM} {Quantum} {Experience}},
  journal = {arXiv preprint arXiv:1805.10928},
  year    = {2018}
}

@book{nielsen2010quantum,
  author    = {Nielsen, Michael A. and Chuang, Isaac L.},
  title     = {Quantum Computation and Quantum Information},
  publisher = {Cambridge University Press},
  year      = {2010}
}

@article{held1962dynamic,
  author  = {Held, Michael and Karp, Richard M.},
  title   = {A dynamic programming approach to sequencing problems},
  journal = {Journal of the Society for Industrial and Applied Mathematics},
  volume  = {10},
  number  = {1},
  pages   = {196--210},
  year    = {1962}
}

@article{liu_optimized_2021,
	title = {An optimized quantum minimum searching algorithm with sure-success probability and its experiment simulation with {Cirq}},
	volume = {12},
	issn = {1868-5137, 1868-5145},
	url = {https://link.springer.com/10.1007/s12652-020-02840-z},
	doi = {10.1007/s12652-020-02840-z},
	language = {en},
	number = {11},
	urldate = {2024-12-18},
	journal = {Journal of Ambient Intelligence and Humanized Computing},
	author = {Liu, Wenjie and Wu, Qingshan and Shen, Jiahao and Zhao, Jiaojiao and Zidan, Mohammed and Tong, Lian},
	month = nov,
	year = {2021},
	pages = {10425--10434},
}

@article{durr1996quantum,
  title={A quantum algorithm for finding the minimum},
  author={Durr, Christoph and Hoyer, Peter},
  journal={arXiv preprint quant-ph/9607014},
  year={1996}
}

@article{grover1998quantum,
  title={Quantum computers can search rapidly by using almost any transformation},
  author={Grover, Lov K},
  journal={Physical Review Letters},
  volume={80},
  number={19},
  pages={4329},
  year={1998},
  publisher={APS}
}

@article{long2001grover,
  author  = {Long, Gui-Lu},
  title   = {{Grover} algorithm with zero theoretical failure rate},
  journal = {Physical Review A},
  volume  = {64},
  number  = {2},
  pages   = {022307},
  year    = {2001}
}

@article{grover1997quantum,
  author  = {Grover, Lov K.},
  title   = {Quantum mechanics helps in searching for a needle in a haystack},
  journal = {Physical Review Letters},
  volume  = {79},
  number  = {2},
  pages   = {325},
  year    = {1997}
}

@inproceedings{grover_fast_1996,
	address = {Philadelphia, Pennsylvania, United States},
	title = {A fast quantum mechanical algorithm for database search},
	isbn = {978-0-89791-785-8},
	url = {http://portal.acm.org/citation.cfm?doid=237814.237866},
	doi = {10.1145/237814.237866},
	language = {en},
	urldate = {2025-01-15},
	booktitle = {Proceedings of the twenty-eighth annual {ACM} symposium on {Theory} of computing  - {STOC} '96},
	publisher = {ACM Press},
	author = {Grover, Lov K.},
	year = {1996},
	pages = {212--219},
}

@article{toyama_quantum_2013,
	title = {Quantum search with certainty based on modified {Grover} algorithms: optimum choice of parameters},
	volume = {12},
	issn = {1573-1332},
	shorttitle = {Quantum search with certainty based on modified {Grover} algorithms},
	url = {https://doi.org/10.1007/s11128-012-0498-0},
	doi = {10.1007/s11128-012-0498-0},
	language = {en},
	number = {5},
	urldate = {2025-01-17},
	journal = {Quantum Information Processing},
	author = {Toyama, F. M. and van Dijk, W. and Nogami, Y.},
	month = may,
	year = {2013},
	pages = {1897--1914},
}

@article{toyama2008multiphase,
  author  = {Toyama, F. M. and Van Dijk, W. and Nogami, Y. and Tabuchi, M. and Kimura, Y.},
  title   = {Multiphase matching in the {Grover} algorithm},
  journal = {Physical Review A},
  volume  = {77},
  number  = {4},
  pages   = {042324},
  year    = {2008}
}

@article{sprung_scattering_1993,
	title = {Scattering by a finite periodic potential},
	volume = {61},
	issn = {0002-9505, 1943-2909},
	url = {https://pubs.aip.org/ajp/article/61/12/1118/1054250/Scattering-by-a-finite-periodic-potential},
	doi = {10.1119/1.17306},
	language = {en},
	number = {12},
	urldate = {2025-01-20},
	journal = {American Journal of Physics},
	author = {Sprung, D. W. L. and Wu, Hua and Martorell, J.},
	month = dec,
	year = {1993},
	pages = {1118--1124},
}

@article{flood_traveling-salesman_1956,
	title = {The {Traveling}-{Salesman} {Problem}},
	volume = {4},
	issn = {0030-364X},
	url = {https://www.jstor.org/stable/167517},
	number = {1},
	urldate = {2025-02-17},
	journal = {Operations Research},
	author = {Flood, Merrill M.},
	year = {1956},
	note = {Publisher: INFORMS},
	pages = {61--75},
}
